\documentclass{cernepprep}

\usepackage{times}

\usepackage{graphicx}% Include figure files
\usepackage{multirow}

\usepackage{textcomp} % need only for 0/00 and micro \textmu
\usepackage{color}

\usepackage{cite}

\usepackage[
colorlinks,
citecolor=blue,
]{hyperref}

\begin{document}

\begin{titlepage}

\EXPnumber{DIRAC/PS212}
\EPnumber{2022-058}
\EPdate{\today}
%\EPdate{XX November 2019}

\title{Investigation of $K^+K^-$ pairs in the effective mass region near $2m_K$}

\begin{Authlist}

B.~Adeva\Iref{s}, 
L.~Afanasyev\Iref{d}, 
A.~Anania\Iref{im},
S.~Aogaki\Iref{b},
A.~Benelli\Iref{cz}, 
V.~Brekhovskikh\Iref{p},  
T.~Cechak\Iref{cz}, 
M.~Chiba\Iref{jt}, 
P.~Chliapnikov\Iref{p},  
D.~Drijard\Iref{c}, 
A.~Dudarev\Iref{d},  
D.~Dumitriu\Iref{b}, 
P.~Federicova\Iref{cz}, 
%D.~Fluerasu\Iref{b}, 
A.~Gorin\Iref{p}, 
%O.~Gorchakov\Iref{d},
K.~Gritsay\Iref{d}, 
C.~Guaraldo\Iref{if}, 
M.~Gugiu\Iref{b}, 
M.~Hansroul\Iref{c}, 
Z.~Hons\Iref{czr}, 
S.~Horikawa\Iref{zu},
Y.~Iwashita\Iref{jk},
V.~Karpukhin\Iref{d}, 
J.~Kluson\Iref{cz}, 
M.~Kobayashi\Iref{k}, 
%V.~Kruglov\Iref{d}, 
L.~Kruglova\Iref{d}, 
A.~Kulikov\Iref{d}, 
E.~Kulish\Iref{d},
%A.~Kuptsov\Iref{d}, 
A.~Lamberto\Iref{im}, 
A.~Lanaro\Iref{u},
R.~Lednicky\Iref{cza}, 
C.~Mari\~nas\Iref{s},
J.~Martincik\Iref{cz},
L.~Nemenov\IIref{d}{c}{*},
M.~Nikitin\Iref{d}, 
K.~Okada\Iref{jks}, 
V.~Olchevskii\Iref{d},
%V.~Ovsiannikov\Iref{v},
M.~Pentia\Iref{b}, 
A.~Penzo\Iref{it}, 
M.~Plo\Iref{s}, 
P.~Prusa\Iref{cz},  
G.~Rappazzo\Iref{im}, 
A.~Romero Vidal\Iref{s},
A.~Ryazantsev\Iref{p},
V.~Rykalin\Iref{p},
J.~Saborido\Iref{s}, 
J.~Schacher\Iref{be},
%J.~Schacher\IAref{be}{*},
A.~Sidorov\Iref{p}, 
J.~Smolik\Iref{cz}, 
F.~Takeutchi\Iref{jks}, 
T.~Trojek\Iref{cz}, 
S.~Trusov\Iref{m}, 
T.~Urban\Iref{cz},
T.~Vrba\Iref{cz},
V.~Yazkov\IAref{m}{\dag}, 
Y.~Yoshimura\Iref{k}, 
%M.~Zhabitsky\Iref{d}, 
P.~Zrelov\Iref{d} 

\end{Authlist}

\Instfoot{s}{Santiago de Compostela University, Spain}
\Instfoot{d}{JINR Dubna, Russia}
\Instfoot{im}{Messina University, Messina, Italy}
\Instfoot{b}{IFIN-HH, National Institute for Physics and Nuclear Engineering, Bucharest, Romania}
\Instfoot{cz}{Czech Technical University in Prague, Czech Republic}
\Instfoot{p}{IHEP Protvino, Russia}
\Instfoot{jt}{Tokyo Metropolitan University, Japan}
\Instfoot{c}{CERN, Geneva, Switzerland}
\Instfoot{if}{INFN, Laboratori Nazionali di Frascati, Frascati, Italy}
\Instfoot{czr}{Nuclear Physics Institute ASCR, Rez, Czech Republic}
\Instfoot{zu}{Zurich University, Switzerland}
\Instfoot{jk}{Kyoto University, Kyoto, Japan}
\Instfoot{k}{KEK, Tsukuba, Japan}
\Instfoot{u}{University of Wisconsin, Madison, USA} 
\Instfoot{cza}{Institute of Physics ASCR, Prague, Czech Republic}
\Instfoot{jks}{Kyoto Sangyo University, Kyoto, Japan}
%\Instfoot{v}{Voronezh State University, Russia}
\Instfoot{it}{INFN, Sezione di Trieste, Trieste, Italy}
\Instfoot{be}{Albert Einstein Center for Fundamental Physics, Laboratory of High Energy Physics, Bern, Switzerland}
%\Instfoot{ba}{Basel University, Switzerland}
\Instfoot{m}{Skobeltsin Institute for Nuclear Physics of Moscow State University, Moscow, Russia}

\Anotfoot{*}{Corresponding author.} 
\Anotfoot{}{$\!\!\!\!^\dag$deceased} 

\Collaboration{DIRAC Collaboration}
\ShortAuthor{DIRAC Collaboration}

\newpage

\begin{abstract}
%{\color{red}{
The DIRAC experiment at CERN investigated in the reaction 
$\rm{p}(24~\rm{GeV}/c) + Ni$ the particle pairs 
$K^+K^-, \pi^+ \pi^-$ and $p \bar{p}$ with relative momentum $Q$ 
in the pair system less than 100~MeV/c. 
Because of background influence studies, DIRAC explored 
three subsamples of $K^+K^-$ pairs, obtained 
by subtracting -- using time-of-flight (TOF) technique -- background 
from initial $Q$ distributions with $K^+K^-$ sample fractions more 
than 70\%, 50\% and 30\%. The corresponding pair distributions 
in $Q$ and in its longitudinal projection $Q_L$ were 
analyzed first in a Coulomb model, which takes into account 
only Coulomb final state interaction (FSI) and assuming 
point-like pair production. This Coulomb model analysis 
leads to a $K^+K^-$ yield increase of about four at $Q_L=0.5$~MeV/c 
compared to 100~MeV/c. In order to study contributions 
from strong interaction, a second more sophisticated model 
was applied, considering besides Coulomb FSI also strong FSI 
via the resonances $f_0(980)$ and $a_0(980)$ and  
a variable distance $r^*$ between the produced $K$ mesons. 
This analysis was based on three different parameter sets 
for the pair production. For the 70\% subsample and 
with best parameters, $3680\pm 370$ $K^+K^-$ pairs was found 
to be compared to $3900\pm 410$ $K^+K^-$ extracted 
by means of the Coulomb model.
%Half of these pairs lie 
%in the mass interval $2M_K$ to $2M_K + 0.8~\rm{MeV}$ 
%($M_K$ is charged kaon mass). 
Knowing the efficiency of the TOF cut
for background suppression, the total number of 
detected $K^+K^-$ pairs was evaluated to be 
around $40000\pm 10\%$, which agrees with the result 
from the 30\% subsample.
The $K^+K^-$ pair number in the 50\% subsample differs from the two other values by about three standard deviations, confirming --- as discussed in the paper --- that experimental data in this subsample is less reliable.
\\
In summary, the upgraded DIRAC experiment observed increased $K^+K^-$ 
production at small relative momentum $Q$. The pair distribution in $Q$ is 
well described by Coulomb FSI, whereas a potential influence from strong 
interaction in this $Q$ region is insignificant within experimental errors.
%}
\end{abstract}
\vspace{2cm}
\Submitted{(To be submitted)}
\end{titlepage}

\section{Introduction}
\label{sec:intro}

The production of oppositely charged meson pairs 
with low relative momentum allows to study Coulomb and 
strong interactions between the two particles 
\cite{URET61, BILE69, NEME85, AFAN96, AFAN93, AFAN94, ADEV04, ADEV05, ADEV11,
ADEV15, ADEV09, ADEV14, ADEV16, ADEV17, LEDN08, LEDN09}. 
In the case of $\pi\pi$ and $\pi K$ free pair investigation, also 
the numbers of generated bound states were evaluated. Furthermore, 
$\pi\pi$ and $\pi K$ atom lifetimes were measured and corresponding 
scattering lengths derived \cite{ADEV11,ADEV17}. Pions and kaons exhibit 
the simplest hadron structure consisting of only two quarks. Therefore, 
$\pi\pi$ and $\pi K$ scattering near threshold is well described by 
low-energy QCD, i.e. chiral perturbation theory (ChPT), 
nonperturbative lattice QCD (LQCD) and dispersion relation analysis. 

The physical properties of $K^+K^-$ Coulomb pairs -- prompt pairs with 
$Q$ distribution enhanced at small $Q$ mainly by Coulomb FSI -- 
and $K^+K^-$ atoms (kaonium) differ from the same properties of 
the $\pi\pi$ and $\pi K$ systems, because strong interaction in 
the $K^+K^-$ system with low relative momentum is affected 
by the presence of the two scalar resonances $f_0(980)$ and 
$a_0(980)$ with masses near $2M_K$. Potential $K^+K^-$ atoms, 
taking into account only Coulomb interaction, show a Bohr radius of 
$r_B$ = 110 fm, a Bohr momentum of $p_B$ = 1.8 MeV/c and a binding energy in 
the ground state of -6.6 keV. These values are not significantly 
changed by strong $K^+K^-$ interaction, because this interaction 
according to \cite{Krewald04} shifts the binding energy only 
by about 3\%. The Coulomb final state interaction has 
a significant influence on the distribution of $Q$, 
the relative momentum in the $K^+K^-$ centre-of-mass system (c.m.). 
The pair production is strongly enhanced with decreasing $Q$. 
This effect is large in the $Q$ region below few $p_B$. Further, 
the kaonium lifetime in the ground state has been calculated under 
different assumptions \cite{Wycech93,Krewald04,Yin-Jie06,Klevansky11}  
%[Wycech93, Krewald04, Zhang06, Klevansky11]
resulting in values in the interval $\tau=(1 - 3) \cdot 10^{-18}$~s. 
This lifetime range is three orders of magnitude smaller than the lifetimes 
of $\pi\pi$ and $\pi K$ atoms. Assuming a lifetime for kaonium in 
the ground state of $\tau \sim 10^{-18}$~s, the produced atoms will 
decay and thus have no time to interact with 
other target atoms and to break up the generating $K^+K^-$ pairs. 
At BNL \cite{Wiencke92}, $10.2\pm3.8$ $K^+K^-$ Coulomb pairs 
were detected. 

In the two data taking runs with similar experimental conditions and with the closed number of proton-Ni interactions (data sets DATA1 and DATA2), DIRAC identified 
about 11000 $K^+K^-$ pairs (30\% subsample). Half of these pairs lie 
in the effective mass interval $2M_K$ to $2M_K + 0.8~\rm{MeV}$. 
The pair distributions in $Q$ and their projections 
were analyzed in order to study the influence of $K^+K^-$ Coulomb 
and strong FSI interaction as well as of the distance $r^*$ 
between the produced K mesons.

\section{Setup and experimental conditions} %2
\label{sec:setup}

\begin{figure}[ht]
\begin{center}
\includegraphics[width=0.85\columnwidth]{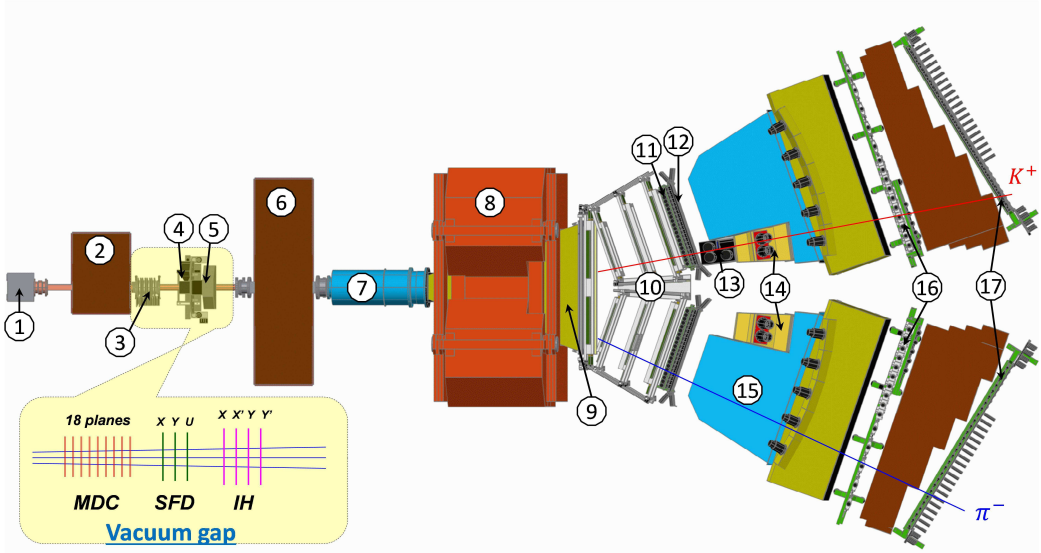}
\caption{ General view of the DIRAC setup 
(1 -- target station;
2 -- first shielding;
3 -- micro drift chambers (MDC);
4 -- scintillating fiber detector (SFD); 
5 -- ionization hodoscope (IH); 
6 -- second shielding; 
7 -- vacuum tube; 
8 -- spectrometer magnet; 
9 -- vacuum chamber; 
10 -- drift chambers (DC); 
11 -- vertical hodoscope (VH); 
12 -- horizontal hodoscope (HH); 
13 -- aerogel Cherenkov (ChA); 
14 -- heavy gas Cherenkov (ChF); 
15 -- nitrogen Cherenkov (ChN); 
16 -- preshower (PSh); 
17 -- muon detector (Mu).}
\label{fig:det}
\end{center}
\end{figure}

The aim of the magnetic 2-arm vacuum spectrometer 
\cite{DIRAC2,GORC05a, GORC05b, PENT15} 
(Fig.~\ref{fig:det}) is to detect and identify  
$K^+ K^-$ , $\pi^+ \pi^-$ , $\pi^- K^+$, and $\pi^+ K^-$ 
pairs with small Q \cite{ADEV17}. The structure of 
$K^+ K^-$ and $\pi^+ \pi^-$  pairs after the magnet is 
approximately symmetric. The 24~GeV/$c$ primary proton beam, 
extracted from the CERN PS, hit a Ni target of 
$(108\pm 1)$~\textmu{}m thickness ($7.4\cdot10^{-3} X_0$). 
With a spill duration of 450~ms, the beam intensity was 
$(1.05 \div 1.2)\cdot10^{11}$ protons/spill, and the 
corresponding flux in the secondary channel $(5 \div 6) \cdot 10^6$ 
particles/spill. 

After the target station, primary protons pass under the setup 
to the beam dump. The axis of the secondary channel is inclined 
relative to the proton beam by $5.7^\circ$ upward. The solid 
angle of the channel is $\Omega = 1.2 \cdot 10^{-3}$ sr. 
Secondary particles propagate mainly in vacuum up to the Al foil 
$(7.6 \cdot 10^{-3} X_{0})$ at the exit of the vacuum chamber, 
which is installed between the poles of the dipole magnet 
($B_{max}$ = 1.65~T and $BL$ = 2.2~Tm).
In the vacuum channel gap, 18 planes of the Micro Drift Chambers (MDC) 
and ($X$, $Y$, $U$) planes of the Scintillation Fiber Detector (SFD) 
were installed in order to measure both the particle coordinates  
($\sigma_{SFDx} = \sigma_{SFDy} = 60$~\textmu{}m, 
$\sigma_{SFDu} = 120$~\textmu{}m) and the particle time 
($\sigma_{tSFDx} = 380$~ps, $\sigma_{tSFDy} = \sigma_{tSFDu} = 520$~ps).  
The total matter radiation thickness between target and vacuum chamber 
amounts to $7.7 \cdot 10^{-2} X_{0}$. Each spectrometer arm is 
equipped with the following subdetectors \cite{DIRAC2}: 
drift chambers (DC) to measure particle coordinates with 
about 85~\textmu{}m precision and to evaluate the particle path length; 
vertical hodoscope (VH) to determine particle times with 110~ps accuracy 
for identification of equal mass pairs via the time-of-flight (TOF) 
between SFDx plane and VH hodoscope; horizontal hodoscope (HH) 
to select in the two arms particles with a vertical distance 
less than 75~mm ($Q_{Y}$ less than 15~MeV/$c$); 
aerogel Cherenkov counter (ChA) to distinguish kaons from protons; 
heavy gas ($\text{C}_{4} \text{F}_{10}$) Cherenkov counter (ChF) to distinguish 
pions from kaons and protons; nitrogen Cherenkov (ChN) and 
preshower (PSh) detector to identify $e^+e^-$; iron absorber and  
two-layer scintillation counter (Mu) to identify muons. 
In the ``negative'' arm, no aerogel counter was installed, 
because the number of antiprotons is small compared to $K^-$. 

Pairs of oppositely charged time-correlated particles (prompt pairs) 
and accidentals in the time interval $\pm$20~ns are selected by 
requiring a 2-arm coincidence (ChN in anticoincidence) with 
the coplanarity restriction (HH) in the first-level trigger. 
The second-level trigger selects events with at least one track 
in each arm by exploiting the DC-wire information (track finder). 
Particle pairs $\pi^{-} p$ ($\pi^{+} \bar{p}$) from 
$\Lambda$ ($\bar{\Lambda}$) decay were used for spectrometer calibration 
and $\text{e}^+\text{e}^-$ pairs for general detector calibration. 

%\newpage

\section{Fractions of $K^+K^-$ pairs with $K^+$ or $K^-$ mesons from the 
resonance decays}
\label{sec:resonance}

To study the $K^+K^-$ Coulomb and strong FSI one has to take 
into account a non point-like $K^+K^-$ pair production. Thus, if one 
hadron of the pair is a decay product of a relatively narrow resonance, 
the relative separation $r^*$ of the hadron production points may be 
substantially increased by the resonance path length $l^*$ in the pair 
c.m.s., which coincides at small $Q$ with the resonance path length in 
the rest frame of the decay hadron $l^* = p_D/(m_h\Gamma)$, where $p_D$ 
is the decay momentum of a hadron of mass $m_h$ and $\Gamma$ is the 
resonance width \cite{Lednic92}. The path lengths of relatively 
narrow resonances such as $K^*(892)$ , $\Lambda(1520)$ and $\varphi(1020)$, 
are in $K^+K^-$ c.m.s. $2.3 fm$, $6.2 fm$ and $11.9 fm$, respectively. 
They should be compared with 
%{\color{red}{
$\langle r^* \rangle = (4/\pi) r_0 \approx 4.5 fm$
%}} 
corresponding to a typical Gaussian radius $r_0 \approx 2 fm$, 
characterizing the $K^+K^-$ correlation function at moderate $Q$-values in 
$pA$ collisions, and the Bohr radius $r_B=110 fm$. 
One may conclude that only the $\varphi(1020)$ path length substantially 
exceeds a typical $r^*$ separation.

Obviously, the increased separation due to the substantial resonance path 
length leads to a weaker Coulomb correlation than in the case of point-like 
pair production. 

%(To introduce the short comment about $r^*$ influence on the strong 
%interaction! Richard!).

%{\color{red}{
In order to take this into account, it is necessary to know the fractions of 
$K^+K^-$  pairs with $K^+$ or $K^-$ such resonance decays. 
The fractions of $K^+K^-$ pairs with $K^+$ or $K^-$ from the decays of 
$K(892)$, $\Lambda(1520)$ and $\phi(1020)$ were determined in \cite{DN2018} 
using the data on $K^+K^-$ pair production and cross sections of
$K(892)$, $\Lambda(1520)$ and $\phi(1020)$
generation in pp interactions at 24GeV/c and 400GeV/c.
%}}

%\begin{equation}\label{eq:sh1}
%\sigma(K^+K^-)=(0.500 \pm 0.028)~{\rm mb}. 
%\end{equation}

Other numerous resonances, some of which are observed only in the phase-shift 
analyses, either have large widths or small branching ratios into the final 
states with kaons and/or small production rates (such as $f_1(1285)$ with
$\Gamma \approx 24 MeV/c^2$ and $Br(K\bar K) \approx 9\%$ or $f’(1525)$ 
with $\Gamma\approx 73 MeV/c^2$ and $Br(K\bar K) \approx 89\%$). 
The contribution of these resonances and direct $K^+K^-$ pairs to the 
distribution on $r^*$ will be described by a Gaussian.

The contributions of $K^*(892)$, $\Lambda(1520)$ and $\phi(1020)$ in $K^+K^-$ 
pairs production were evaluated as the product of the branching with 
generation of charged $K$ meson and the relative value of the dedicated 
inclusive cross section. Following \cite{DN2018}  the relative contribution 
of all types of $K^*(892)$ equals to:

\begin{equation}\label{fraction-from-k*}
f_{K^*} (K^+K^-) = (45 \pm 10)\%
\end{equation}

The fraction of $K^+K^-$ pairs with the $K^-$ from the $\Lambda(1520)$ 
decay amounts to:

\begin{equation}\label{fraction-from-lambda}
f_{\Lambda(1520)}(K^+K^-) = (8 \pm 2)\%
\end{equation} 

The $K^+K^-$ pair from one and the same $\phi$ decay doesn't contribute to 
$K^+K^-$ pairs at small $Q$. The contribution of $K$ meson from $\phi$ 
decay in the interval of small $Q$ is possible when $\phi$
is associated at least with a pair of strange particles (dominantly kaons).  
The  cross sections of associated $\phi$ production measured at $24 GeV/c$ 
and $400 GeV/c$ are quite different, which may result from a bad kaon
identification in the bubble chamber experiment at $24 GeV/c$ and expected
increase of the associated production with increasing energy, thus leading to
a conservative estimate \cite{DN2018}: 

\begin{equation}\label{fraction-phi-kk}
f_\phi(K^+K^-)=(2-14)\%
\end{equation}

The errors in the $f$ values do not include the uncertainty 
of the approach used in \cite{DN2018}. 
Therefore, in the following we estimate the finite-size FSI effect on
the $K^+K^-$ yield and $Q$ spectrum taking into account, besides a Gaussian
short-distance contribution, also the ones containing exponential tails
due to kaons from the decays of $K^*(890)$, $\Lambda(1520)$ and
$\phi(1020)$ resonances using the fractions 
(\ref{fraction-from-k*})-(\ref{fraction-phi-kk}) to construct $r^*$
distributions with minimum and maximum values of average $r^*$.

\section{Production of free $K^+K^-$ pairs} %4
\label{sec:freebound}

As mentioned in section \ref{sec:resonance} the prompt $K^+K^-$ pairs, emerging 
from proton-nucleus collisions, are produced manly from 
short-lived sources. These pairs undergo Coulomb and strong FSI 
resulting in modified unbound states (Coulomb pair) or forming bound systems. 
The accidental pairs arise from different proton-nucleus interactions.

\subsection{Point-like $K^+K^-$ production and Coulomb FSI}  %4.1

Taking into account the Coulomb FSI only (Fig.\,\ref{fig:r_dist}\,(a)), the 
production of unbound oppositely charged $K^+K^-$ pairs from short-lived 
sources, i.e. Coulomb pairs, is described \cite{NEME85} in the point-like 
production approximation, by
\vspace{-2mm}
\begin{equation}\label{eq:cross_sect_C}
\frac{d^6\sigma_C}{d^3\vec p_{K^+} d^3\vec p_{K^-}} =
\frac{d^6\sigma^0_s}{d^3\vec p_{K^+} d^3\vec p_{K^-}} A_C(q) 
\quad \mbox{with} \quad
A_C(q) = \frac{2\pi m_K \alpha/q}
{1-\exp\left( -2\pi m_K \alpha/q\right) } \;.
\end{equation}
\vspace{-2mm}

where $\vec p_{K^+}$ and $\vec p_{K^-}$ are the momenta of the charged kaons,
$\sigma_s^0$ is the inclusive production cross section of $K^+K^-$ pairs 
from short-lived sources without FSI and the Coulomb enhancement function 
$A_C(q)$ represents the non-relativistic $K^+K^-$ Coulomb wave function 
squared at zero separation, well-known as the
Gamov-Sommerfeld-Sakharov factor \cite{GAMO28, SOMM31, SAKH91}.

\subsection{Non point-like $K^+K^-$ production and strong and Coulomb FSI} %4.2 

\begin{figure}[ht]
\hspace{-17mm}
\begin{tabular}{p{40mm}p{120mm}}
\begin{center}
\vspace{2mm}
\includegraphics[width=16mm]{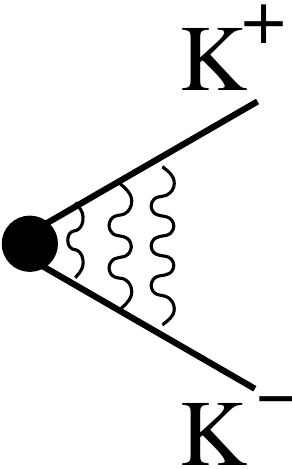} 
\end{center}
&
\begin{center}
\includegraphics[width=130mm]{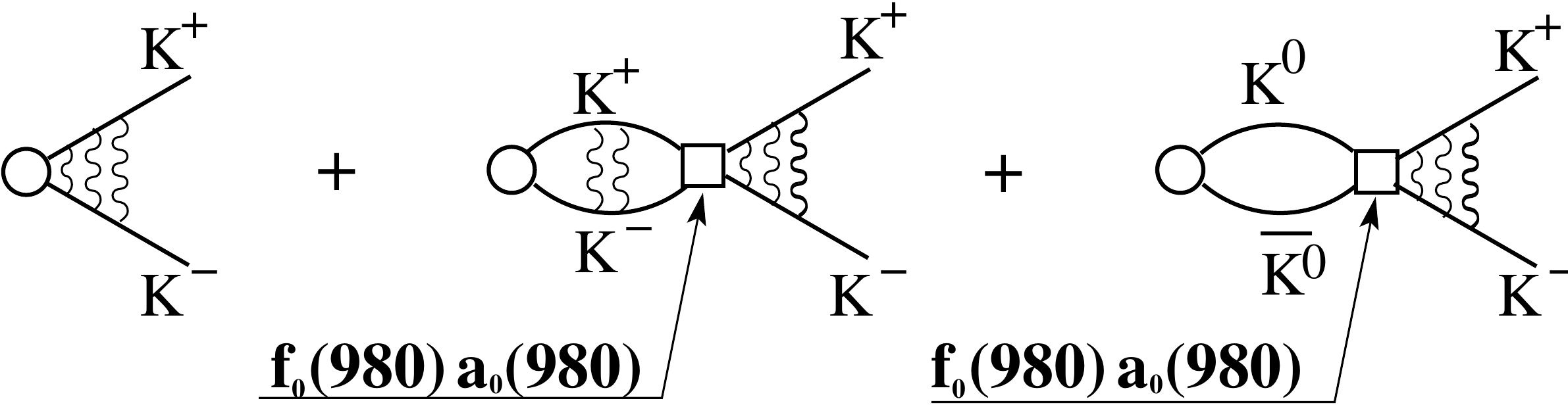}
\end{center}
\\[-21pt]
\begin{center}
(a)
\end{center}
&
\begin{center}
(b)
\end{center}
\end{tabular}
\vspace{-2mm}
\caption{
The schematic description of $K^+K^-$ production processes.
a) The black point presents the pair point-like production; 
the wavy lines describe the Coulomb interaction in the final state. 
b) The circle and square present the pair non point-like production and 
strong interaction in the final state respectively.
}
\label{fig:r_dist}
\end{figure}

Up to now, the production of $K^+K^-$ pairs (\ref{eq:cross_sect_C}), was 
assumed to be point-like and only the Coulomb FSI was taken into account. 
The influence of the finite size effects and hadron strong interaction in 
the final state on the production of free and bound $K^+K^-$ pairs 
(Fig.\,\ref{fig:r_dist}\,(b)), was considered in \cite{LEDN08,LEDN09,Lednic82} 
and used to fit experimental $K^+K^-$ correlation functions in experiments 
NA49 \cite{NA49}, STAR \cite{STAR} and ALICE \cite{ALICE}.

As for the $K^+K^-$ strong interaction near threshold, it is dominated
by the spin-0 isoscalar ($T = 0$) and isovector ($T = 1$) resonances
$f^0(980)$ and $a^0(980)$ characterized by their masses $M_r$ and
respective couplings $\gamma_r$ - to the $K {\bar K}$ channel and
$\gamma_r\,'$ - to the $\pi\pi$ and $\pi\eta$ channels for $f^0(980)$ and
$a^0(980)$, respectively \cite{Lednic82}, \cite{Bekele07}, \cite{ALICE},
\cite{Martin}, \cite{Achasov}, \cite{Antonelli}.

There is a great deal of uncertainty in the properties of these
resonances reflected in uncertainties of their PDG widths: $10-100 MeV$
and $50-100$ MeV for $f^0(980)$ and $a^0(980)$, respectively. Fortunately,
the dominant imaginary parts of the scattering lengths are basically
determined by the ratios $\gamma_r / \gamma_r\,'$ with rather small
uncertainty. As for the real parts of the scattering lengths, due to the 
closeness of $f^0$ and $a^0$ masses to the $K \bar K$ threshold, they are 
quite uncertain and rather small, varying in existing fits from 
$-0.3 fm$ to $0.3 fm$.

To calculate the $K^+K^-$ correlation function, we use the $f^0(980)$
and $a^0(980)$ parameters from Martin et al.\cite{Martin}, Achasov et al.
\cite{Achasov} and ALICE \cite{ALICE}. The ALICE parameters for $a^0(980)$
coincide with those from Achasov et al., and, for $f^0(980)$, they are
determined from a fit of the ALICE $K^+K^-$ correlation functions.

Note that the ALICE $K^+K^-$ correlation data \cite{ALICE} disagrees 
with the $f^0(980)$ parameterisations from Martin et al.\cite{Martin}, 
Achasov et al.\cite{Achasov}, and Antonelli \cite{Antonelli}. 
The ALICE $K_s\,K^\pm$
correlation data \cite{ALICEKsK} (with the absent $f^0$ contribution)
excludes $a^0(980)$ parameters from Martin et al., favouring those from
Achasov et al., while the STAR and ALICE $K_s\,K_s$ correlation data 
\cite{Bekele07,ALICEKsKs}, 
is unable to discriminate among all these parameterizations.
\vspace{-3mm}

%=======================================================================

\section{Data processing} %5
\label{sec:dp}

The collected events were analyzed with the DIRAC reconstruction program 
ARIANE \cite{Ariane} modified for analyzing $K K$ data. 

\subsection{Tracking} %5.1
\label{ssec:track}

Only events with one or two particle tracks in DC   
of each arm are processed. The event reconstruction is 
performed according to the following steps \cite{ADEV17}:
\begin{itemize}
\item One or two hadron tracks are identified in DC  
of each arm with hits in VH, HH and PSh slabs and 
no signal in ChN and Mu.
\item Track segments, reconstructed in DC, are extrapolated 
backward to the beam position in the target, 
using the transfer function of the dipole magnet and 
the program ARIANE. This procedure 
provides approximate particle momenta and  
the corresponding points of intersection in MDC, SFD and IH.
\item Hits are searched for around the expected SFD coordinates 
in the region $\pm 1$~cm corresponding to (3--5)~$\sigma_\text{pos}$ 
defined by the position accuracy taking into account 
the particle momenta. The number of hits around the two tracks is 
$\le 4$ in each SFD plane and  $\le 9$ in all three SFD planes. 
In some cases only one hit in the region $\pm 1$~cm occurred.  
To identify the event when two particles crossed the same SFD column 
was requested the double ionisation in the corresponding IH slab. 
\end{itemize}
The momentum of the positively or negatively charged particle 
is refined to match the $X$-coordinates of the DC tracks 
as well as the SFD hits in the $X$- or $U$-plane, 
depending on the presence of hits. In order to find 
the best 2-track combination, the two tracks may not use 
a common SFD hit in the case of more than one hit in the proper region. 
In the final analysis, the combination with the best $\chi^2$ 
in the other SFD planes is kept. 

\subsection{Setup tuning using $\Lambda$ and $\bar{\Lambda}$ particles} %5.2
\label{ssec:geom}

In order to check the general geometry of the DIRAC experiment, 
the $\Lambda$ and $\bar{\Lambda}$ particles, decaying into 
${\rm p}\pi^-$ and $\pi^+\bar{\rm p}$ in our setup, were 
used \cite{ADEV17}. After setup tuning
the weighted average value of the experimental $\Lambda$ mass over all runs, 
$M^\mathrm{DIRAC}_{\Lambda} = (1.115680 \pm 2.9 \cdot 10^{-6})$~GeV/$c^2$, 
agrees very well with the PDG value,  
$M^\mathrm{PDG}_{\Lambda} = (1.115683 \pm 6 \cdot 10^{-6})$~GeV/$c^2$. 
The weighted average of the experimental ${\bar{\Lambda}}$ mass is 
$M^\mathrm{DIRAC}_{\bar{\Lambda}} = (1.11566 \pm 1 \cdot 10^{-5})$~GeV/$c^2$. 
This demonstrates that the geometry of the DIRAC setup is well described. 

The width of the $\Lambda$ mass distribution allows to test the momentum and 
angular setup resolution in the simulation. Table~\ref{tab:lambda-w} shows 
a good agreement between simulated and experimental $\Lambda$ width in 
DATA1 and DATA2.
A further test consists in comparing the experimental $\Lambda$ and 
${\bar{\Lambda}}$ widths. 

\begin{table}[htbp]
\caption{$\Lambda$ width in  GeV/$c^2$  for experimental and MC data and ${\bar{\Lambda}}$ width for experimental data.}
\label{tab:lambda-w}
\begin{center}
\begin{tabular}{|c|c|c|c|}
\hline 
\rule{0pt}{2.5ex} &   $\Lambda$ width (data)  &  $\Lambda$ width (MC)  & ${\bar{\Lambda}}$ width (data) \\
&      GeV/$c^2$     &     GeV/$c^2$   &     GeV/$c^2$ \\
\hline   
%\rule{0pt}{2.5ex}  DATA1 &    $4.33 \cdot 10^{-4} \pm 8.2 \cdot 10^{-6}$ &  $ 4.38 \cdot 10^{-4} \pm 4.6 \cdot 10^{-6}$  &  $4.6 \cdot 10^{-4} \pm 2 \cdot 10^{-5}$ \\
%\hline   
\rule{0pt}{2.5ex}  DATA1 &    $4.42 \cdot 10^{-4} \pm 7.4 \cdot 10^{-6}$ &  $ 4.42 \cdot 10^{-4} \pm 4.4 \cdot 10^{-6}$  &  $4.5 \cdot 10^{-4} \pm 3 \cdot 10^{-5}$ \\
\hline   
\rule{0pt}{2.5ex}  DATA2 &    $4.41 \cdot 10^{-4} \pm 7.5 \cdot 10^{-6}$ &  $ 4.37 \cdot 10^{-4} \pm 4.5 \cdot 10^{-6}$  &  $4.3 \cdot 10^{-4} \pm 2 \cdot 10^{-5}$ \\
\hline
\end{tabular}
\end{center}
\end{table}

The average value of correction which was introduced in the simulated
width is $1.00203\pm0.00191\cdot10^{-3}$. 
%{\color{red}{
This number to be used for 
the introduction of the non significant corrections in the l.s. particle's 
momenta.
%}}
 
%{\color{red}{
The $Q_L$ distribution of 
$\pi^+\pi^-$ pairs can be used to check the geometrical alignment. 
Since the $\pi^+\pi^-$ system is symmetric, the corresponding $Q_L$ 
distribution should be centered at 0. 
The experimental $Q_L$ distribution of pion pairs with transverse momenta 
$Q_T < 4$~MeV/$c$, is centered at 0 with a precision of 0.2~MeV/$c$.
%}}

%\begin{figure}[h]
%\begin{center}
%\includegraphics[width=0.6\columnwidth]{Fig-4.jpg}
%\end{center}
%\caption{$Q_L$ distribution of $\pi^+\pi^-$ experimental data (DATA1 to DATA2). }
%\label{fig:6_3}
%\end{figure} 

\subsection{Event selection} %5.3
\label{ssec:Ev_sel}

%{\color{red}{
The processed events were collected in DATA1 and DATA2. 
%}}
Equal mass pairs contained in the selected event 
sample are classified into three categories: $K^+K^-$, $\pi^+\pi^-$ and
$p\bar{p}$ pairs.   

The classification is based on the TOF measurement \cite{note2001}. 
In the momentum range from 3.8 to 7~$\text{GeV}/c$, 
additional information from the Heavy Gas Cherenkov (ChF) counters 
(Section~\ref{sec:setup}) is used to better separate $\pi^+\pi^-$ from 
$K^+K^-$ and $p\bar{p}$ pairs. The ChF counters detect pions in this region 
with (95--97)\% efficiency \cite{note1305}, whereas kaons and protons 
(antiprotons) do not generate any signal. 
%{\color{red}{
Due to the finite resolution of  
the TOF system and the Cherenkov efficiency, the selected $K^+K^-$ sample 
with high momentum pairs 
still contains about 10\% $\pi^+\pi^-$ and 10\% $p\bar{p}$ events. 
%}}

\begin{figure}[h]
\begin{center}
\includegraphics[width=0.6\columnwidth]{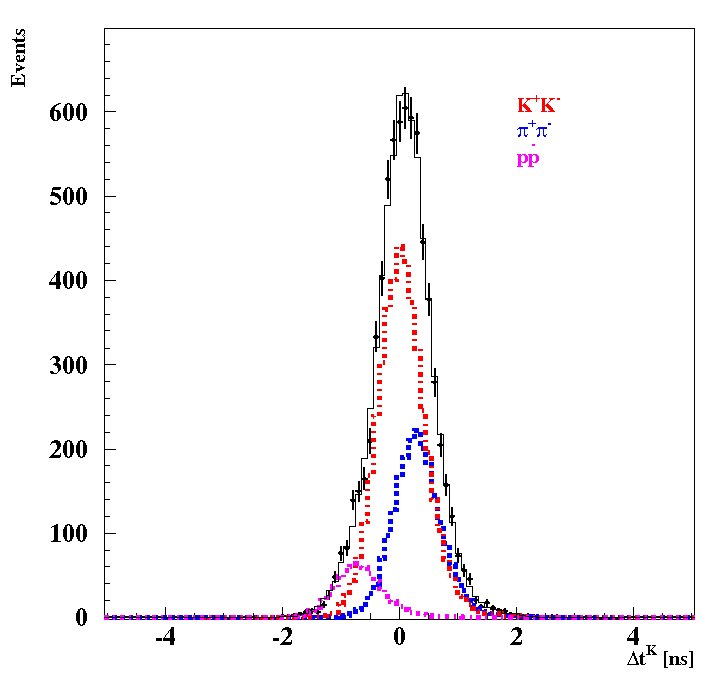}
\end{center}
\vspace{-7mm}
\caption{
$\Delta t^{K}$ distribution of  $K^+K^-$, $\pi^+\pi^-$ and 
$p\bar{p}$ pairs with momentum of about 3.5$\text{GeV}/c$. 
The peak at zero corresponds to $K^+K^-$, the peak on the right to 
$\pi^+\pi^-$ and the small peak on the left to $p\bar{p}$ pairs. 
}
\label{fig:5_3_1}
\end{figure}

The TOF is measured and calculated for the distance between the SFD X-plane 
and the VH of about 11m. The length and momentum of each track are evaluated 
using the tracking system. The relative precision of the momentum measurement 
is about $3\times 10^{-3}$. For 'positive' and 'negative' tracks, the 
expected TOF $t^{calc}_\pm$ is calculated assuming that it is $K^+K^-$ pair. 
Furthermore, the difference between calculated and measured TOF, 
$\Delta t_{\pm}=t^{calc}_{\pm} - t^{exp}_{\pm}$, was determined. 
In order to classify the pairs, the averaged difference 
$\Delta t=\frac{1}{2}(\Delta t_{+}+\Delta t_{-})$ was used. 
The  $\Delta t^{K}$ distribution of events corresponding to 
a momentum of about 3.5$\text{GeV}/c$ is presented in 
Fig.~\ref{fig:5_3_1}.  

\begin{figure}[h]
\begin{center}
\includegraphics[width=0.55\columnwidth]{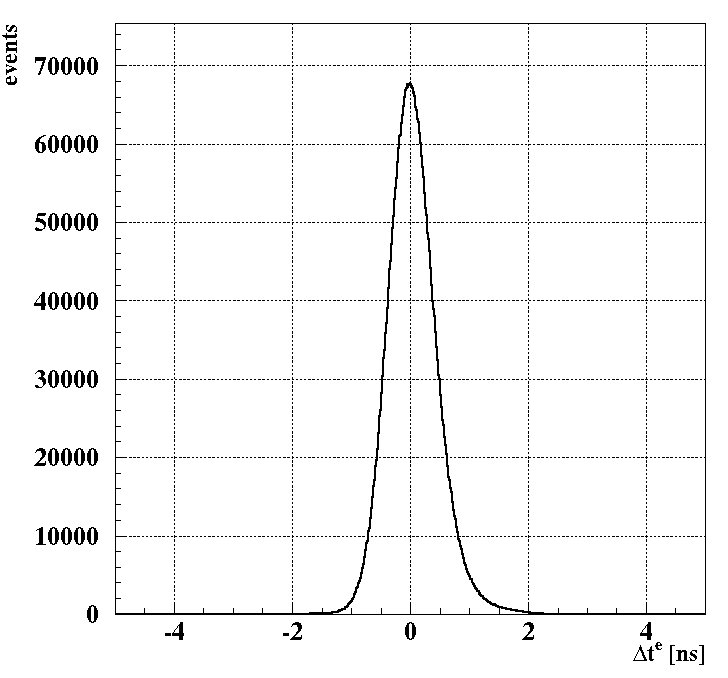}
\end{center}
\vspace{-7mm}
\caption{$\Delta t^{e}$ distribution of electron-positron pairs.}
\label{fig:5_3_2}
\end{figure}

To evaluate the amount of pairs in each category, model distributions of 
$\Delta t^{K}$ obtained from $e^+e^-$ pairs are used \cite{note2001}. 
These $e^+e^-$ data were collected for calibration purposes with 
a dedicated trigger (Section~\ref{sec:setup}) during standard data taking. 
Again, the average difference $\Delta t^{e}$  between expected and 
measured TOF for the electron and positron was calculated assuming 
electron mass. The $\Delta t^{e}$ distribution shown in Fig.~\ref{fig:5_3_2} 
exhibits a half width at half maximum of 440~ps corresponding to 
the time resolution of the TOF system. 

The $\Delta t^{K}$ distributions of $K^+K^-$, $\pi^+\pi^-$ and $p\bar{p}$ pairs 
at fixed lab momentum $p_{lab}$ show the same shape as for $e^+e^-$. 
The $K^+K^-$ peak is at zero, whereas the $\pi^+\pi^-$ and $p\bar{p}$ peaks 
are on the positive and negative side, respectively. The distance 
of the $\pi^+\pi^-$ and $p\bar{p}$ peak from zero is increasing with 
decreasing $p_{lab}$. 

The experimental data are spread over a wide momentum interval (2.5--7)~GeV/$c$. 
The shape of the $\Delta t^{K}$ distribution depends on momentum and 
on its interval width. Therefore, the data are analyzed within bins of 
a new variable $\Delta T_{K-\pi}$. For each track in a pair, 
the $\Delta t$ parameter was calculated in two versions: 
1) using kaon mass ($\Delta t^{K}$) and 
2) using pion mass ($\Delta t^{\pi}$). 
The new parameter $\Delta T_{K-\pi}$ is then defined as the difference 
between the TOFs calculated for kaon and pion (for each pair track): 
\begin{equation}\label{eq:deltat_k_p} 
\Delta T_{K-\pi}=\frac{1}{2}(\Delta T^+_{K-\pi} + \Delta T^-_{K-\pi}) = 
\Delta t^{K} - \Delta t^{\pi}. 
\end{equation}
In the analysis, the data are processed in one hundred 25~ps wide 
$\Delta T_{K-\pi}$ bins. 

The advantage of this technique is the constant shape of 
the $\Delta t^{K}$ distribution of $\pi^+\pi^-$ and $K^+K^-$ pairs 
for different $\Delta T_{K-\pi}$ values. The selection of 
a particular $\Delta T_{K-\pi}$ bin fixes the distance between 
the peak positions of the distributions corresponding to $K^+K^-$ and 
$\pi^+\pi^-$ pairs. The distance $\Delta T_{K-\pi}$ between 
the peaks of the $K^+K^-$, $\pi^+\pi^-$ and $p\bar{p}$ pairs is maximal 
for pairs with minimal momentum $p^{min}_{lab}=2.5$GeV/$c$. 

The model distributions of  $\pi^+\pi^-$, $K^+K^-$ and $p\bar{p}$ pairs  
are used to fit the experimental distributions. In each 25~ps 
$\Delta T_{K-\pi}$ bin, the amount of events is determined for 
the three categories as shown in Fig.~\ref{fig:5_3_1}. 
The collected data consists mainly of $\pi^+\pi^-$ pairs. 
For analyzing $K^+K^-$ pairs, subsets with a significant $K^+K^-$ portion 
are needed. In each $\Delta T_{K-\pi}$ bin, contiguous bins in 
$\Delta t^{K}$ are selected by demanding the $K^+K^-$ population 
to exceed a certain threshold. Hence, we consider three subsamples of events 
containing at least a $K^+K^-$ population of 30\%, 50\% and 70\%. 
The cleanest so-called 70\% $K^+K^-$ sample consists of only 
$K^+K^-$ pairs with high momenta, where Cherenkov counters suppress 
$\pi^+\pi^-$ pairs efficiently. 

% This part corresponds to line_245 to line_261 in kk-210530.pdf! 
\section{Experimental results} %6.
\label{sec:Exp-res}
				
For DATA1 and DATA2, the $K^+K^-$, $\pi^+\pi^-$ and 
$p\bar{p}$ pair numbers were evaluated in the 30\%, 50\% and 
70\% subsample (Table \ref{tab:p2}). The number of proton interaction 
with the target in DATA1 and DATA2 are nearly the same. 

\begin{table}[h]
  \caption{
Pair numbers in DATA1 and DATA2, evaluated in 
the three subsamples (30\%, 50\%, 70\%). The $R$ is the ratio of events in correspondent subsample to the full number (all). 
}
\label{tab:p2}
  \centering		
\setlength\extrarowheight{3pt}
    \begin{tabular}{|c|c|c|c|c|c|c|c|}
\hline 
DATA1 &\multicolumn{4}{c|}{Experimental data ($Q_T < 15~\rm{MeV}/c$)} 
& \multicolumn{3}{c|}{$R$ (\%)} 
\\ \hline \hline
Sample & all &  30\% &  50\% & 70\% & 30\%/all & 50\%/all & 70\%/all 
\\[3pt] \hline
$\pi^+\pi^-$ & $7.77 \cdot 10^{6}$ & 17290 & 3540 &  620 & 0.22 & 0.05 & 0.008 
\\[3pt] \hline
$K^+K^-$ & 90840 & 25660 & 15040 & 8210 & 28.2 & 16.6 & 9.0 
\\[3pt] \hline
$p\bar p$ & 7670 & 2960 & 1930 & 880 & 38.6 & 25.2 & 11.5 
\\[3pt] \hline
\end{tabular} 
\\[12pt]

\begin{tabular}{|c|c|c|c|c|c|c|c|}
\hline 
DATA2 &\multicolumn{4}{c|}{Experimental data ($Q_T < 15~\rm{MeV}/c$)} 
& \multicolumn{3}{c|}{{$R$ (\%)}} 
\\ \hline \hline
Sample & all &  30\% &  50\% & 70\% & 30\%/all & 50\%/all & 70\%/all 
\\ \hline
$\pi^+\pi^-$ & $7.96 \cdot 10^{6}$ & 15230 & 2970 &  80 & 0.19 & 0.04 & 0.001
\\ \hline
$K^+K^-$ & 92960 & 25550 & 15910 & 8330 & 27.5 & 17.1 & 9.0 
\\ \hline
$p\bar p$ & 7200 & 2950 & 1780 & 770 & 41.0 & 24.7 & 10.7 
\\ \hline 
\end{tabular} 
\end{table}   

%{\color{red}{ 
It can be seen that the number of $K^+K^-$ pairs in DATA1 and DATA2  
without cutting in the three subsamples are consistent. 
%In the right part of Table \ref{tab:p2}, the number of pairs in the 
%subsample divided on the total pairs number (residuals R) are presented in \%.
%}}
The experimental data were obtained with 
a trigger restriction on $Q_T$ at about 15~MeV/$c$. 
For the final analysis, data were used with the software restriction 
$Q_T < 6$~MeV/$c$, where the setup efficiency is constant. 
To study a possible influence of the $Q_T$ limit on 
$R$, a larger data sample with $Q_T < 8$~MeV/$c$ was also analyzed. 
The resulting pair numbers for $Q_T < 8$~MeV/$c$ are decreased by 1.8 and 
the corresponding $R$ values in agreement with those in Table \ref{tab:p2}. 
 
All events in the three samples are prompt.
Their numbers and distributions on any parameter were 
evaluated by subtracting the background of the accidental events
using the time difference between VH hodoscopes. 
The percentage of accidentals before subtraction in the  70\%, 50\% 
and 30\% samples was 9.6\%, 22\% and 47\%, respectively. The 70\% sample 
is the most reliable for the $K^+ K^-$ pair analysis, because 
%{\color{red}{
the total background of accidentals, $\pi^+ \pi^-$ and $p \bar p$ 
prompt pairs is significantly smaller than in the two other samples. 
After background subtraction, the $K^+ K^-$ purity is the highest one. 
%}}

\subsection{The simulation procedure}

The experimental distributions of $K^+K^-$ pairs were compared with the 
corresponding simulated spectra according to different theoretical models. 
The simulated $K^+K^-$ spectra in the pair c.m.s. were calculated
using the relation:
\vspace{-7mm}

\begin{equation}\label{delta-n}
\frac{dN}{dQ_i}=|M_{prod}|^2 F(Q_i) F_{corr}(Q_i)
\end{equation}
\vspace{-5mm}

where $Q_i$ is $Q$ or $Q_L$, $M_{prod}$ the production matrix element without 
the $Q$ dependence in the investigated $Q$ interval, $F(Qi)$ the phase space 
and $F_{corr}(Q_i)$ the correlation function. This function takes into 
account the Coulomb FSI in the Coulomb approximation ($A_c(Q)$) or the 
Coulomb and strong FSI in the more precise models. For the c.m.s. pair is 
added the l.s. momentum $\vec P_{lab}$ that allows to calculate the 
$\vec P^+$ and $\vec P^-$ momentua of the $K^+$ and $K^-$ in l.s and 
their total momentum $\vec P = \vec P^+ + \vec P^-$.

By means of the dedicated code GEANT-DIRAC, the simulated pairs are 
propagated through the setup, taking into account multiple scattering, 
the response of the detectors before the magnet on the $K^+K^-$ pairs
and the response of the detectors after magnet on the single particle. 
Using the information from the detectors the events were reconstructed 
by the code ARIANE and processed as experimental pairs. Then, their
$Q_L$ and $Q$ distributions were calculated and compared with the 
corresponding experimental spectra. The $\vec P_{lab}$ distribution 
was obtained by requiring that $\vec P$ spectrum must fit the experimental 
$K^+K^-$ pair spectrum in $\vec P_{exp}= \vec P^+_{exp} + \vec P^-_{exp}$
where $\vec P^+_{exp}$ and $\vec P^-_{exp}$ are experimental l.s. momentum 
of $K^+$ and $K^-$.

\subsection{Analysis of $Q_L$ and $Q$ distributions}
\label{sec:analysis}
\vspace{-2mm}

\begin{figure}[h]
\begin{minipage}{150mm}
\begin{center}
\includegraphics[width=0.55\columnwidth]{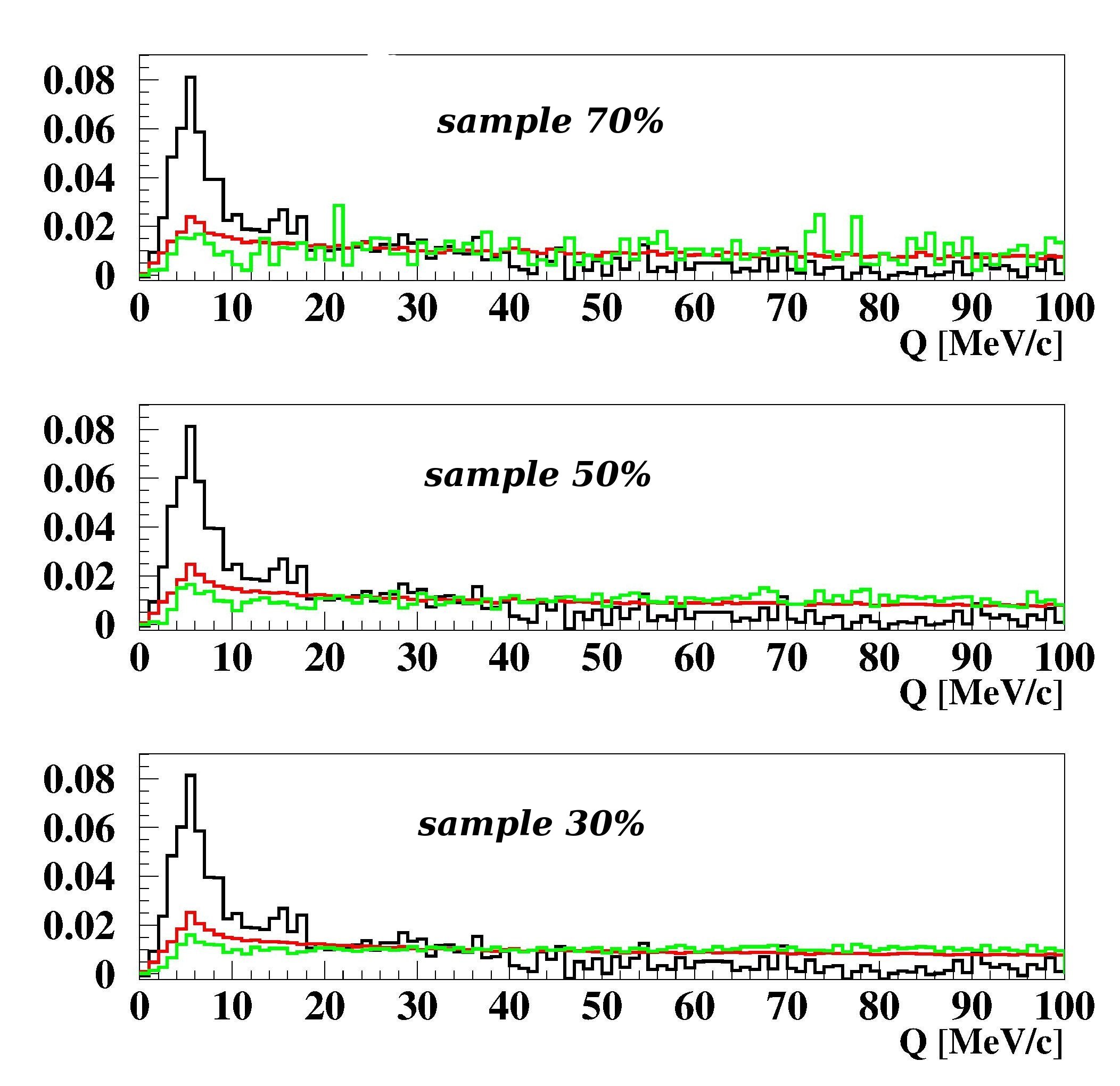} %MP
\end{center}
\end{minipage}
\vspace{-1mm}
\caption{The $Q$ distribution of $\pi^+\pi^-$ (red), $p\bar p$ (black) and
accidentals (green) pairs calculated as $K^+K^-$ pairs.
} 
\label{fig:back-all}
\end{figure}

The subtraction of $\pi^+\pi^-$, $p\bar p$ and accidentals background is 
based on the estimated time and momentum setup resolutions. 
A statistical fluctuation and possible systematic uncertainty of this 
subtraction may lead to a residual background, distorting the $K^+K^-$ 
distributions in $Q$ and $Q_{L}$. The fractions of $K^+K^-$ and residual 
background pairs can be evaluated using different shapes of their $Q$ 
and $Q_{L}$ distributions.

The distributions of accidentals and $\pi^+\pi^-$ pairs were obtained 
calculating these experimental pairs as $K^+K^-$ system for each subsample. 
Due to small yield of $p\bar p$ pairs only one sample containing bins with 
their population greater than $50\%$ was produced. The $p\bar p$ sample 
was processed as $K^+K^-$ system and used for all three subsamples analysis. 
The $Q$ spectra of the three background types for all subsamples are shown 
in Fig.\ref{fig:back-all}.

Taking into account a small difference between the shapes of $\pi^+\pi^-$ 
and accidental background distributions, we fit the $K^+K^-$ and residual 
background fractions, assuming the same shape of $\pi^+\pi^-$ and
accidental background $Q$ and $Q_{L}$ distributions, i.e., considering 
the $\pi^+\pi^-$ and $p\bar p$ background only.

The effect and background values must not depend from the distribution 
type chosen. To check it the dedicated analysis was done for $Q$ and 
$Q_{L}$ experimental distributions using fitting curve and only $\pi^+\pi^-$ 
and $p\bar p$ background. In this analysis the $K^+K^-$ pairs distribution 
in $Q$, $Q_{L}$ is calculated proposing the point-like $K^+K^-$ pairs 
production with only Coulomb interaction in the final state (Coulomb 
parametrization). 
For each run and each subsample, the experimental $Q$, $Q_{L}$ 
distributions $D_{exp}$ were fitted by sum of the three distributions 
according to the formula:

\begin{equation}\label{fit-three-distrib}   
D_{exp} = N_{KK}D_{KK}^n + N_{\pi\pi}D_{\pi\pi}^n 
+ N_{p\bar p}D_{p\bar p}^n
\end{equation}
	  
where $D_{\pi\pi}^n$ and $D_{p\bar p}^n$ denote corresponding background distributions of $\pi^+\pi^-$ and $p\bar p$ pairs normalized to unity, $D_{KK}^n$ is the simulated $K^+K^-$ distribution normalized to unity. $N_{\pi\pi}$ and $N_{p\bar p}$ are free fitted parameters indicating number of $\pi^+\pi^-$ and $p\bar p$ pairs.

The number of $K^+K^-$ pairs, $N_{KK}$, is given by the constraint

\begin{equation}\label{kk-constraint}
N_{exp} = N_{KK} + N_{\pi\pi} + N_{p\bar p}     
\end{equation}

where $N_{exp}$ is the total number of events in given $D_{exp}$
distribution. In this case the errors of the $K^+K^-$ pairs and total 
background number are equal.

For the six distributions on $Q$ and $Q_L$ (DATA1 and DATA2, 
three subsamples) all the $\chi^2/\mathrm{ndf}$ values are within the interval 
0.7-1.2, and the $K^+K^-$ numbers of the two runs in each subsample are 
in agreement. The $\chi^2/\mathrm{ndf}$ values for $Q$ distributions in DATA2/DATA1 
runs are $0.98/1.19, 0.77/1.08$ and $0.69/0.87$ for 70\%, 50\% and 30\% 
subsamples respectively. 
The probability density function (PDF) has maximum
around $\chi^2/\mathrm{ndf}=1$ and decreased by half for $\chi^2/\mathrm{ndf}=0.85$ and $1.15$. The Coulomb parametrization describes all 6 experimental distributions well because the PDF values for 70\% and 50\% subsamples for two data sets are near maximum at $\chi^2/\mathrm{ndf}\approx 1$ and for 30\% subsample this parameter is near maximum for 
DATA2 and for DATA1 it deflects from maximum with 0.31 value, which is acceptable.

\begin{figure}[h]
\vspace{-2mm}
\begin{minipage}{150mm}
\begin{center}
\includegraphics[width=0.60\columnwidth]{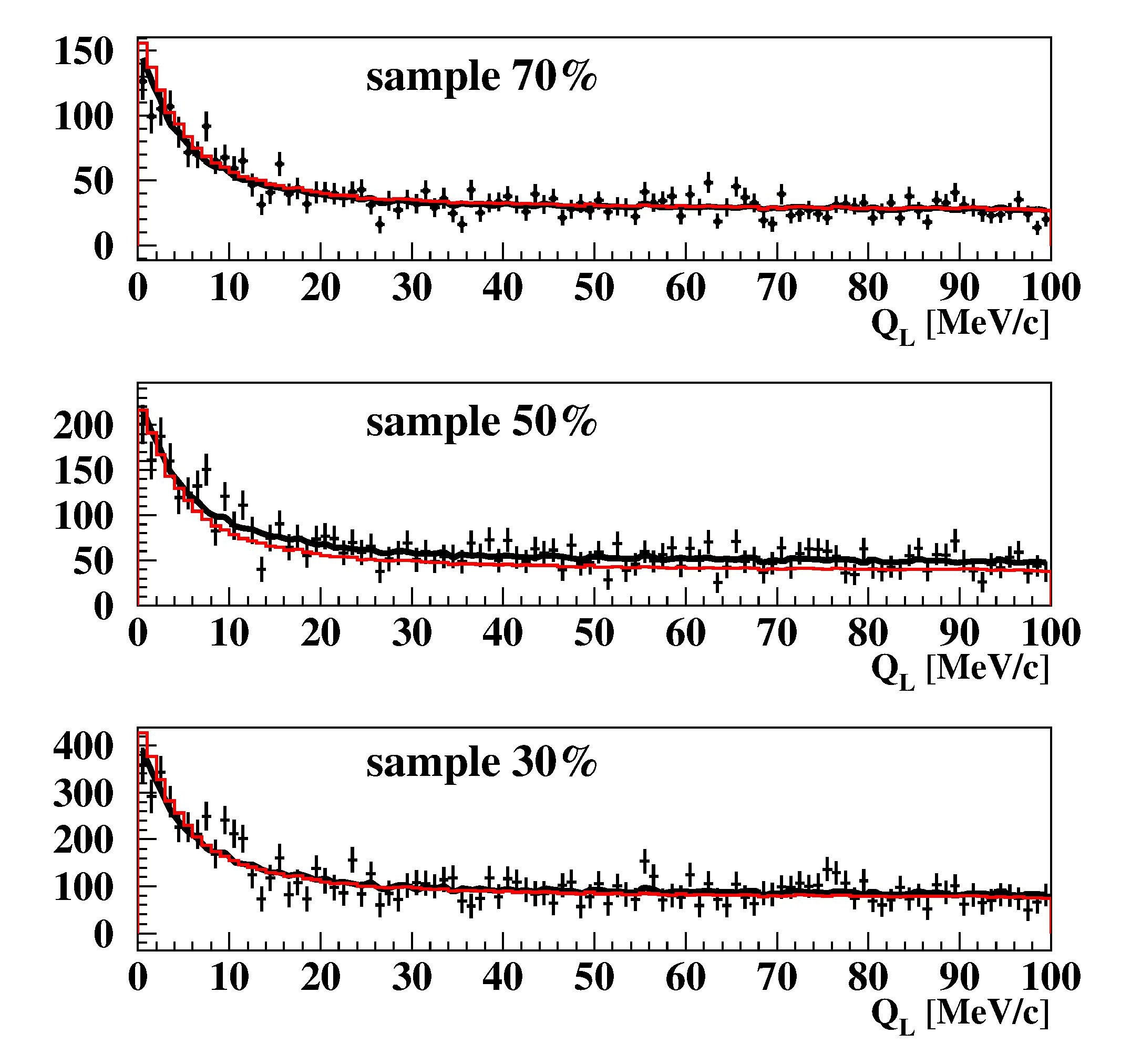} %MP
\end{center}
\end{minipage}
\vspace{-1mm}
\caption{
$Q_L$ distributions of the subsamples 30\%, 50\% and 70\% for DATA1 
and DATA2. The experimental spectra in the interval
$0 < Q_L < 100$ MeV/c are fitted by simulated $K^+K^-$ (point-like
Coulomb FSI) and residual $\pi^+\pi^-$ and $p\bar p$ background distributions.
The red curve is the 
%{\color{red}{
$K^+K^-$ distribution, the black one is the sum of $K^+K^-$ 
%}}
and residual background distributions. In the subsamples 70\% 
and 30\% the residual background is small and these curves practically 
coincide. For $K^+K^-$ pairs in the region of $Q_L < 10$ MeV/c the Coulomb 
enhancement is clearly visible, whereas the residual background is small.
} 
\label{fig:p6-bis}
\end{figure}

Figure \ref{fig:p6-bis}  presents the experimental distributions in $Q_L$, the 
fitting curves for $K^+K^-$ pairs and the sum of the total background and 
fitting distributions.
It is seen that in the 70\% and 30\% subsamples the fitting curves  
coincide practically with the experimental distributions demonstrating 
that the residual background is small. 
In the 50\% subsample the background level is significantly higher 
and the fitting curve is lower than the experimental points. 

A strong enhancement in the pair yield can be recognized in the $Q_L$ 
distributions between 0 to 10 MeV/c. It is caused by the Coulomb final 
state $K^+K^-$ pairs interaction, because the residual background is small. 
The same analysis was performed for $Q$ distributions.

Table \ref{tab:p8} presents the outcome of the two analyses and demonstrates 
a good agreement for the $K^+K^-$ pair numbers obtained in the $Q_L$ and $Q$ 
distribution analysis.

The $K^+K^-$ pair numbers presented in Table \ref{tab:p8} 
were obtained with the residual background description using only 
$\pi^+\pi^-$ and $p\overline p$ pairs. 
The fits, where accidental background was added, to give
the same numbers of $K^+K^-$ pairs within 0-0.2 errors.

\begin{table}[h]
\begin{center}
\begin{minipage}{95mm}
\caption{
Matching pair numbers for $Q$ and $Q_L$ distribution analyses.
The errors of $K^+K^-$ and background values are the same.
}
\label{tab:p8}
\setlength\extrarowheight{5pt}
    \begin{tabular}{|c|c|c|c|p{19mm}|}
\hline 
 & cut on ToF & distribution & $K^+K^-$ &
 $\pi^+\pi^-$ \& $p \bar p$ background \\
\hline \hline
\multirow{6}{*}{\setlength\extrarowheight{0pt}\setlength\tabcolsep{0pt}
\begin{tabular}{c} DATA1\\+\\DATA2 \end{tabular}} 
& \multirow{2}{*}{$70\%$} & $Q$ & $3900\pm 410$ & $-110$ 
\\ \cline{3-5}
& & $Q_L$ & $3930\pm 580$ & $-140$  
\\ \cline{2-5}
& \multirow{2}{*}{$50\%$} & $Q$ & $5320\pm 730$ & $1100$ 
\\ \cline{3-5}
& & $Q_L$ & $5460\pm 1020$ & $960$   
\\ \cline{2-5}
& \multirow{2}{*}{$30\%$} & $Q$ & $11220\pm 1370$ & $180$ 
\\ \cline{3-5}
& & $Q_L$ & $10750\pm 2020$ & $300$ 
\\ \hline 
\end{tabular} 
\end{minipage}
\end{center}
\end{table}

\subsection{Data analysis assuming non point-like $K^+K^-$ pair production 
and Coulomb and strong $K^+K^-$ interaction in the final state}

In section \ref{sec:analysis} the $K^+K^-$ pairs were analyzed 
assuming their point-like production and taking into account only Coulomb 
interaction in the final state. In this section the $K^+K^-$ distributions
in $Q$ will be analyzed taking 
%{\color{red}{
into 
%}}
account non point-like pairs production 
and their Coulomb and strong interactions in the final state. It will use 
three theoretical parametrizations: Achasov et al.\cite{Achasov}, 
Martin et al.\cite{Martin} and ALICE \cite{ALICE}.

The distribution in the distance $r^*$  between two $K$ mesons in the general 
case is presented as the sum of four distributions in $r^*$ connecting 
the $K$ mesons from the decay of short-lived sources and long-lived  
resonances:
\vspace{-5mm}

\begin{equation}\label{resonances}
w_g*Gauss+w_{K^*}*K^*(892) + w_\Lambda *\Lambda(1520)
+ w_{\varphi^*}*\varphi(1020)
\end{equation}

%{\color{red}{
The first term describes the contributions of the short-lived sources 
approximated by Gaussian with the radius $r_0 \approx 1.5 fm$, 
the other terms describe the contributions of the three resonances. 
The $w_i$ are the relative contributions of the different sources in 
$K^+K^-$ pair production. The weights values were evaluated using the 
numbers and their errors presented in the equations (\ref{fraction-from-k*}), 
(\ref{fraction-from-lambda}), (\ref{fraction-phi-kk}) 
and the requirement that the sum of $w_i$ equals unity.
%}}

The analysis was performed for the three sets of $w_i$. The first extreme 
set (0.00, 0.76, 0.10, 0.14) maximizes the contributions of $K^*(892)$, 
$\Lambda(1520)$ and $\varphi(1020)$ resonances producing the largest value 
of average $r^*$; the third extreme set (0.57, 0.35, 0.06, 0.02) maximizes 
the role of the short-lived $K^+K^-$ pairs sources generating the minimum 
value of the average $r^*$ and the second set 
(0.10, 0.76, 0.08, 0.06) is using the intermediate values of $w_i$. 
The $Q$ distribution (fitting curves) were calculated for each of DATA1, 
DATA2 and for each sample, using three theoretical parametrizations of 
Achasov, Martin and ALICE \cite{ALICE,Martin,Achasov}.

The experimental data of the 70\% subsample was analyzed by dedicated fitting 
curve with $\pi^+\pi^-$ and $p\bar p$ background. 
The results obtained are shown in Table \ref{tab:7}.
The background errors are the same as for $K^+K^-$ pairs.
The $K^+K^-$ pairs yield is increasing with enlarging  $r^*$ value. 
The difference between extreme yields values gives the maximum numbers of 
systematic errors in connection with the uncertainty of $r^*$ 
distribution. The errors values are $\pm 70, \pm 55$ and $\pm40$. 
These systematic errors are significantly smaller than the errors in 
Table \ref{tab:7}. Therefore for the analysis of the two other experimental 
subsamples, we will use only the intermediate $r^*$ distribution. 
The results of the 70\%, 50\% and 30\% subsamples are presented in 
Table \ref{tab:7}.

\begin{table}[h]
\caption{Pair numbers in the DATA1 and DATA2. The number of $K^+K^-$ 
and background pairs (in brackets) were evaluated by fitting experimental 
distributions on $Q$ in three subsamples by $K^+K^-$ distributions, 
calculated with different parametrizations. The errors of $K^+K^-$ and 
background pairs are identical.}
\label{tab:7}
\begin{tabular}{||c|c|c|c|p{9mm}||}  
\hline \hline
       &  Achasov &  Martin &  ALICE  & Total
\\
  & $K^+K^-$ (backgr.) & $K^+K^-$ (backgr.) & $K^+K^-$ (backgr.) & events
\\ \hline \hline 
70\% sample  & & & & 3790  
\\ \hline 
maximum $r^*$ & $3190\pm 330$    & $3650\pm 370$  & $3720\pm 380$ & 
\\ \hline 
intermediate $r^*$ & $3120\pm 320$ (670) & $3600\pm 360$ (190) & $3680\pm 370$ (110) &
\\ \hline 
$\chi^2/\mathrm{ndf}$ DATA2/DATA1 & $1.03/1.20$ & $1.00/1.18$ & $1.00/1.18$  &
\\ \hline
minimum $r^*$ & $3050\pm 320$    &  $3540\pm 360$  &  $3640\pm 370$ &
\\ \hline \hline   
50\% sample  & & & & 6420  
\\ \hline 
intermediate $r^*$ & $4340\pm 570$ (2080) & $4940\pm 640$ (1480) & $5040\pm 660$ (1380) & 
\\ \hline 
$\chi^2/\mathrm{ndf}$ DATA2/DATA1 & $0.80/1.04$ & $0.79/1.04$ & $0.78/1.05$ & 
\\ \hline \hline
30\% sample  & & & & 11030  
\\ \hline
intermediate $r^*$ & $9230\pm 1080$ (1800) & $10500\pm 1220$ (530) & $10680\pm 1240$ (350) & 
\\ \hline 
$\chi^2/\mathrm{ndf}$ DATA2/DATA1 & $0.70/0.89$ & $0.68/0.88$ & $0.68/0.88$ & 
\\ \hline \hline 
\end{tabular}
\end{table}

\begin{figure}[h]
\centering
\includegraphics[width=0.45\columnwidth]{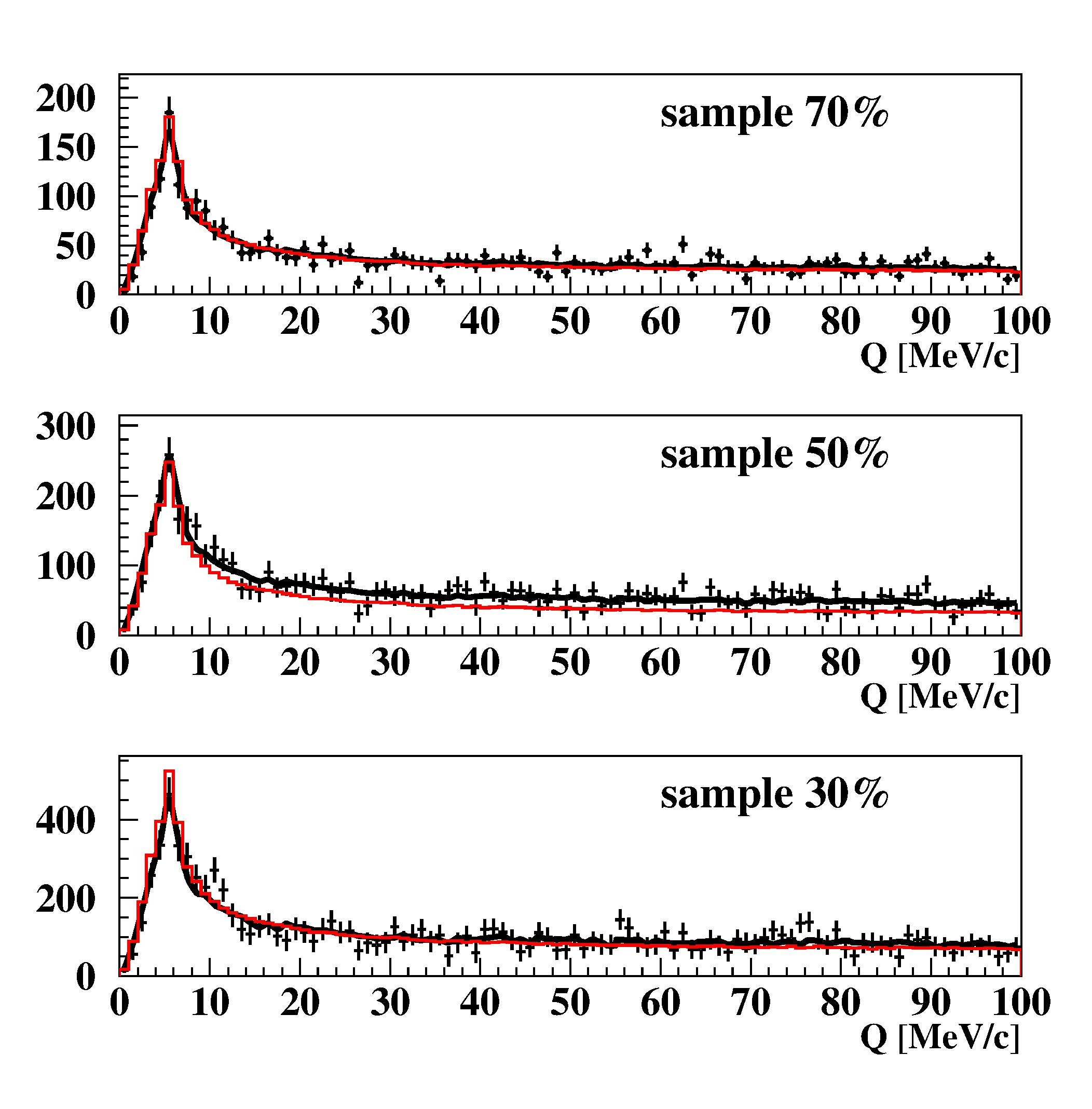}
\vspace{-4mm}
\caption{
$Q$ distributions of the subsamples 30\%, 50\% and 70\% for 
DATA1 and DATA2. Simulated distributions of $K^+K^-$ (ALICE parametrization) 
and 
%{\color{red}{
residual background of $\pi^+\pi^-$, accidental and $p\bar p$ pairs are 
fitting the 
%}}
experimental spectrum in the interval 
%{\color{red}{
$0 < Q < 100~MeV/c$. 
%}}
The red line is the $K^+K^-$ distribution, the black line is the sum of 
$K^+K^-$ and residual background. In the subsamples 70\% and 30\% the 
residual background is small and these lines practically coincide. 
}
\label{fig:pp9}
\end{figure} 

It is seen from Table \ref{tab:7} that for any subsample the Achasov 
parametrization gives the residual background deflection from zero 
significantly larger than Martin and ALICE calculations. The large level 
of residual background can be considered as a result of insufficient accuracy 
of the fitting curve describing $K^+K^-$ distribution on $Q$. The additional 
reason for the better precision of Martin and ALICE parametrization can be 
obtained from the residual background estimation. The expected numbers of 
$\pi^+\pi^-$, $p\bar p$ and accidental pairs in 70\%, 50\% and 30\% 
subsamples are $1050 \pm 50$, $5300 \pm 120$ and $30370 \pm 630$ 
respectively. The errors include the systematical and statistical  accuracy 
of the expected background level evaluation and background statistical 
fluctuations. After expected background subtraction the real residual 
background can differ from zero to one-three errors.

In Achasov parametrization in the 70\% subsample the background deflection 
is 13 standard deviations. In the same subsample the respective deviations 
for Martin and ALICE parametrizations are 3.8 (2.2) standard deviations. 
Therefore in the present paper the experimental data will be 
analyzed using as a main ALICE and Martin parametrizations. 

The large residual background in the 50\% subsample indicates that this 
experimental distribution is less reliable than the 70\% subsample which 
will be used for the calculation of the $K^+K^-$ pairs total number.
The $\chi^2/\mathrm{ndf}$ values of Martin and ALICE parametrizations for 70\% and 
50\% subsamples are slightly better than the same quantities  obtained using 
Coulomb parametrization. It shows that experimental data precision is not
enough to choose, using $\chi^2/\mathrm{ndf}$ values, between simple description 
using point-like production and only Coulomb FSI and more precise 
theoretical approaches taking into account non point-like pairs production,
Coulomb and strong FSI. In the future analysis the Martin and 
ALICE results  obtained with a more accurate theoretically approach will 
be used.

%(To present later the influence on the KK number
%the adding to the background accidentals!). 

Figure \ref{fig:pp9} shows the experimental distributions in $Q$, 
fitting curves describing $K^+K^-$ pairs (ALICE parametrization) and sum 
of fitting curves and residual backgrounds. It is seen that for 70\% (30\%) 
subsample the fitting curve alone describes well the experimental 
distribution in the total interval of $Q$ demonstrating that admixture 
of the residual background to the $K^+K^-$ pairs is relatively small. 
This result is in agreement with the average level of residual 
background, it equals 3\% (3.2\%) of the total number of events in the 
distribution. The same analysis was done for the 50\% subsample.

\subsection{Evaluation of the total number of detected $K^+K^-$ pairs}

\begin{figure}[h]
\centering
\includegraphics[width=0.45\columnwidth]{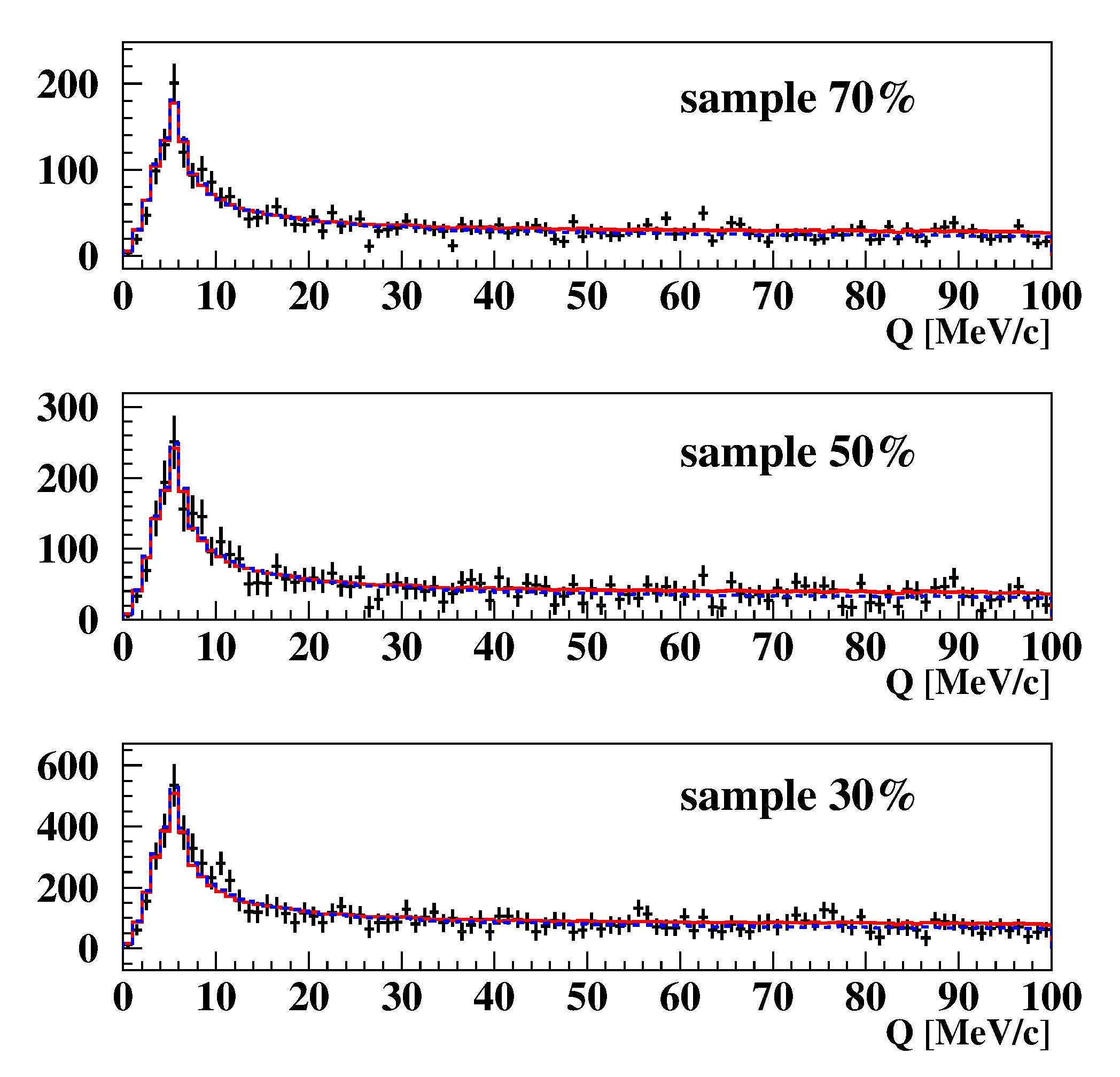}
\vspace{-2mm}
\caption{
$Q$ distributions in the interval $0-100~MeV/c$ of $K^+K^-$ 
(ALICE parametrization) 
of the subsamples 30\%, 50\% and 70\% for the DATA1 and DATA2 after 
residual background subtraction. The red and the blue fitting curves were 
evaluated from the analysis of experimental distributions with residual 
background in Coulomb and Martin parametrizations respectively. It is seen that 
the difference between these curves is not significant and they describe 
"pure" experimental $K^+K^-$ distributions well.
}
\label{fig:p9}
\end{figure} 

\begin{figure}[h]
\centering
\includegraphics[width=0.45\columnwidth]{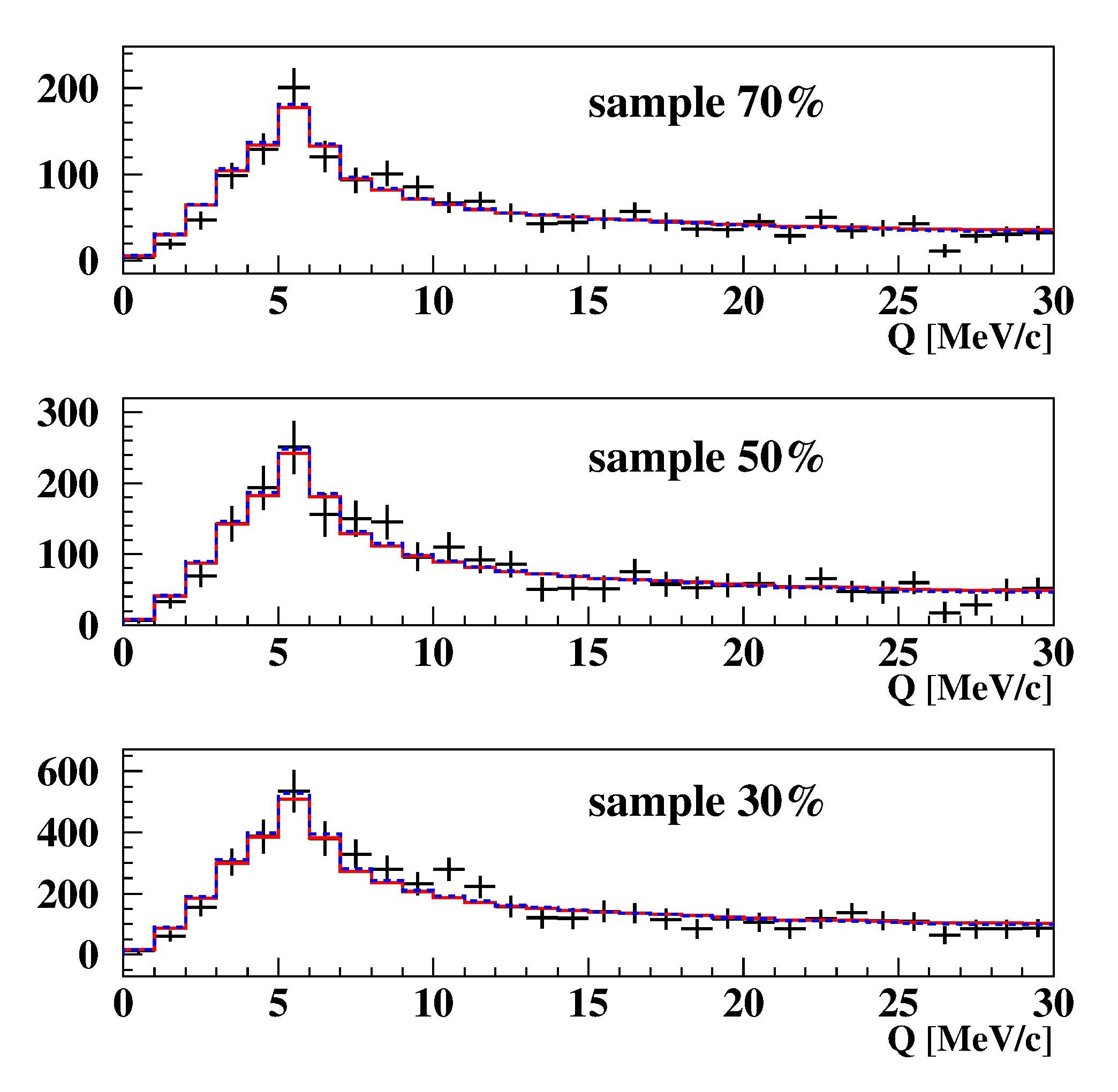}
\vspace{-3mm}
\caption{
$Q$ distributions in the interval $0-30~MeV/c$ of $K^+K^-$ 
(ALICE parametrization) of the subsamples 30\%, 50\% and 70\% for the 
DATA1 and DATA2 after residual background subtraction. The red and 
the blue fitting curves were evaluated from the analysis of experimental 
distributions with residual background in Coulomb and Martin parametrizations 
respectively. It is seen that in this interval of $Q$ the difference between 
these curves is absent and they describe "pure" experimental $K^+K^-$ 
distributions well.
}
\label{fig:pp10}
\end{figure} 

The experimental distributions after residual background subtraction using 
ALICE parametrization are shown  in Figure \ref{fig:p9} together with the 
fitting curves of Martin and point-like Coulomb parametrizations.
The average background level in the corrected experimental distributions 
in the 70\% and 30\% subsamples are less than 3\%. It is seen from the 
numbers of $K^+K^-$ pairs presented in Tables \ref{tab:p8} and \ref{tab:7}  
that the point-like Coulomb parametrization is giving the yield of $K^+K^-$
pairs by 7\%-8\% (5\%-6\%) more than Martin (ALICE) parametrization in the $Q$ 
interval $0-100MeV/c$. The yields difference caused by strong $K^+K^-$ 
interaction in the final state is taken into account only in Martin 
(ALICE) parametrization. The distributions in the $Q$ interval $0-30MeV/c$ 
are presented in Figure \ref{fig:pp10}. It is seen that Martin and Coulomb 
fitting curves describe well the corrected experimental data.

In Table \ref{tab:7} are presented the $K^+K^-$ pairs numbers for 
DATA1 and DATA2. Using the residual $R$ values from Table \ref{tab:p2} 
were calculated the total number of $K^+K^-$ pairs with 
$Q_t<6MeV/c$, $0<Q<100MeV/c$ detected in the experiment. 
It is seen from Table \ref{tab:p2} that the number of $K^+K^-$ pairs 
in the 3 subsamples for DATA1 and DATA2 are in agreement. The total number 
of detected $K^+K^-$ pairs evaluating from the most reliable 70\% subsample
is $40890\pm 4110$. The same values calculated with the 50\% and 30\% 
subsamples are $29650\pm 3880$ and $38140\pm 4430$ pairs respectively.
The $K^+K^-$ pairs number in the 50\% subsample differs from the two other 
values by about 3 standard deviations, confirming as mentioned above, 
that experimental data in this subsample is less reliable. The total 
number of $K^+K^-$ pairs calculated using the Martin parametrization 
from the ALICE parametrization values is significantly smaller than the 
presented errors.

The ratio of $K^+K^-$ pairs to the total number of subtracted background 
pairs in the 70\% subsample case is 10 times larger than in the 30\% 
subsample one. Nevertheless the total numbers of $K^+K^-$ pairs are in good 
agreement, demonstrating that the background and residual background 
subtractions were done correctly.

%\newpage
%\clearpage

\section{Conclusion}

The DIRAC experiment at CERN detected in the reaction 
$\rm{p}(24~\rm{GeV}/c) + Ni$ the particle pairs 
$K^+K^-,\pi^+\pi^-$ and $p \bar{p}$ with relative momentum $Q$ 
between $0 - 100~\rm{MeV}/c$. 
The $Q$ spectrum of $K^+K^-$ pairs was studied with the cut 
on the transverse component $Q_T < 6~\rm{MeV}/c$. 
Three subsamples with $Q$ distributions of $K^+K^-$ pairs 
were obtained by subtracting background from initial 
experimental distributions with $K^+K^-$ populations larger than 70\%, 
50\% and 30\%. The $K^+K^-$ pair numbers, including residual 
background pairs, are 3790 (70\%), 6420 (50\%) and 11030 (30\%). 

These pair distributions in $Q$ and its longitudinal projection $Q_L$ 
were analyzed in two theoretical models. In the first model, only 
Coulomb FSI was taken into account, assuming point-like pair production. 
In the second more precise approach, three theoretical models were used, 
which consider Coulomb and strong FSI interactions via the resonances 
$f_0(980)$ and $a_0(980)$ and the dependence of these interactions 
in the distance $r^*$ between the produced $K$ mesons. 

The analysis based on the first model showed, that 
the Coulomb FSI interaction increases the yield of $K^+K^-$ pairs 
about four times at $Q_L = 0.5~\rm{MeV}/c$ compared to 
$100~\rm{MeV}/c$. 

In the second approach, the $K^+K^-$ strong interaction was described 
using three parameter sets obtained in the Achasov, Martin and ALICE studies 
\cite{Achasov,Martin,ALICE} by analyzing the experimental data sets. 
The numbers of $K^+K^-$ and residual background pairs 
were obtained using different $Q$ shapes of 
these pair distributions. In each subsample, 
the experimental spectrum was described as the sum of $K^+K^-$ 
and residual background distributions. It was shown, that 
the Coulomb model, ALICE and Martin parametrizations are describing 
the experimental distributions in the three subsamples well. 
The best description is provided by the set of 
ALICE parameters with the following number of $K^+K^-$ pairs: 
$3680 \pm 370~(70\%), 5040 \pm 660~(50\%)$ and $10680\pm 1240~(30\%)$ 
with a background level of 3\% for the 70\% and 30\% subsamples. 
The same numbers of $K^+K^-$ pairs, evaluated in the first model, are 
$3900 \pm 410, 5320 \pm 730$ and $11220\pm 1370$, 
which differ from the corresponding ALICE values less than one error. 
The shape of the $K^+K^-$ spectrum in the $Q$ interval 
$0 - 30~\rm{MeV}/c$ is nearly the same in the ALICE, Martin and 
Coulomb parametrizations in all subsamples. Also the distribution of the 
distance $r^*$ between the produced kaons does not have a measurable 
effect on the $Q$ spectrum. 

The experimental precision of the DIRAC data does not allow to choose 
between the simple Coulomb model and the more precise Martin and ALICE 
approaches, taking into account Coulomb and strong FSI, because all 
three models give practically the same $\chi^2$ and the levels of the 
residual background were evaluated correctly. 

The total numbers of detected $K^+K^-$ pairs were evaluated in the ALICE 
parametrization using the known cuts on time-of-flight, which were used 
to suppress the background level. This number is $40890 \pm 4110$ for 
the most reliable 70\% subsample and $38140\pm 4430$ for the 30\% subsample. 
The $K^+K^-$ pair number in the 50\% subsample differs from 
the two other values by about three standard deviations, 
confirming -- as mentioned above -- that experimental data in 
this subsample is less reliable. The total $K^+K^-$ pair number 
calculated with the Martin parametrization deviates from 
the ALICE parametrization values less than the presented errors. 

The ratio of $K^+K^-$ number to the total background level in the 70\% 
subsample is 10 times larger than in the 30\% subsample. Nevertheless, 
the total numbers of $K^+K^-$ pairs are in good agreement, demonstrating
that the background and residual background subtractions were done correctly.

\end{document}